# Influence of Skill and Knowledge of Programmers on Program Behavior Visualization by CS Unplugged

Sumika Jinnouchi
*Kindai University*
Higashi-osaka, Japan
2433340477y@kindai.ac.jp

Masateru Tsunoda
*Kindai University*
Higashi-osaka, Japan
tsunoda@info.kindai.ac.jp

*Abstract*—Computer science unplugged (CS unplugged) is a method of teaching computer science and computational thinking. It does not use a computer but employs physical materials. As CS unplugged, past studies proposed a new method to visualize programming behavior, and evaluated the method based on the understanding test of program. However, the past studied did not consider the factors affecting the test such as participants knowledge. This study analyzed the relationship between the factors and the test results.

## I. INTRODUCTION

Recently, computer science and computational thinking education have gained attention [2]. Computational thinking, which consists of skills such as abstraction and algorithmic thinking, improves problem-solving abilities in the real world. Computer science unplugged (CS unplugged) is a method of teaching computer science and computational thinking [1]. For education, a CS unplugged does not use a computer but employs physical materials such as printed paper. Therefore, CS unplugged can be used at a lower cost, and teachers are not required to manage students' computers.

In this study, our focus is on scenarios where CS unplugged is employed as an introductory method to programming. The aim of the application is to illustrate program behavior and to prevent confusion about the program stack through visualization. We assume that beginners in programming are not accustomed to visually imagining a program's status. In other words, beginners do not explicitly recognize which line of code is executed.

Based on the assumption, study [3] proposed a new method to visualize programming behavior, using a ball and pipes. Study [3] did not experimentally evaluate whether the method helps to understand programming behavior or not. Hence, study [4] posed understanding test of programming to participants, and evaluated the effect of the method. However, the study [4] did not analyze the influence of other factors such as knowledge of participants. This study focused on the factors, and analyze the relationships between factors and the effect, to validate the effect of the method.

## II. VISUALIZING PROGRAM BEHAVIOR

Study [3] proposed how to denote program behavior using a toy which consists of pipes, panes, and balls. The panels are used to join pipes, and the ball moves on the pipes. Such toy is sold by various companies such as [5].

Figure 1 shows an example of visualized program behavior based on the method [3]. The method premises the following:

- One or more pipes corresponds to each line of source code.
- The position of the ball denotes a line which is currently executed. That is, the ball plays a role of the program counter.
- When the ball is goes out of a course which is made by pipes, the program is regarded to end.

Generally, when program ends, outputs of the program such as sound and displayed images are vanished. Therefore, the method also premises that when the ball is goes out of a course (i.e., the program ends), such outputs are also vanished. To sustain the outputs, the method visually explains that a looped course to keep the boll moving (i.e., the program executing) is needed.

## III. EXPERIMENT

**Overview**: We evaluated the visualization method proposed in study [3]. The participants of the experiment were seven undergraduate students who major in computer science. Program execution process was explained by three different ways. After each explanation, the participants answered questions which confirm whether they understand the program behavior correctly or not. Therefore, the participants answered the questions three times. The three explanation ways are the followings:

1. A movie explains that a program loop is needed to prevent stopping program execution.
2. A movie explains movement of a ball on pipes which makes a program loop in detail.

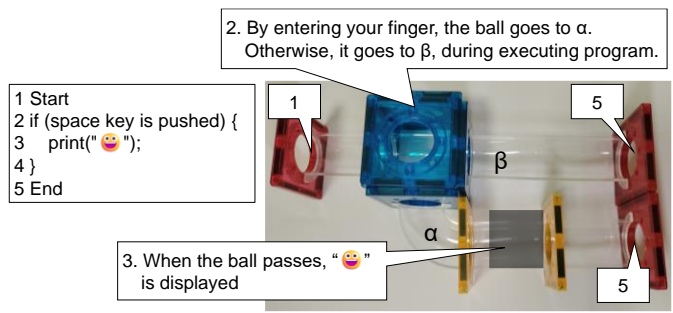

Fig. 1  **Visualization of conditional branch behavior on question (ii)**

3. Using actual ball and pipes, we explained why program loop is needed to prevent stopping execution in person.

Explanation 1 and 3 showed an example which illustrates that proper program loop is needed to hear sound of a bell (i.e., sound output). Explanation 2 was performed, because we assumed that explanation 1 is insufficient for participants to comprehend the behavior of a program loop.

**Comprehension test**: We asked participants to answer two questions about program comprehension. On question (i), line 2 and 4 of Figure 1 were deleted, and asked which type of loop should be added to keep image output on line 3. On the question, we showed two types of loops as shown in Figure 2. Correct answers to question (i) are type A and B. On question (ii), which type should be added to the program shown in Figure 1. This program has to wait users' input, and hence the answer to the question is type B.

Answers to the question (i) and (ii) were collected using Google Forms. Note that on the forms, we also showed visualized program when type A or B is added, as shown in Figure 3. Therefore, participants need not to picture structure of each program denoted by pipes, but to imagine movement of a ball on the program.

**Programming skill and knowledge**; We assumed that programming skill and knowledge affected the comprehension test of programming explained above. Hence, we asked participants to answer the following inquiries on a 10-point Likert scale.

- I1. Do you feel programming is not easy?
- I2. Do you recognize the need for a loop to receive users' input?
- I3. Do you recognize the need for a loop to keep sound?

## IV. Result

**Effect of visualization method**: Table I shows the number and percentage of correct answers to question (i) and (ii). After explanation 3, the percentage was 71%, and it was higher than explanation 1. Therefore, the explanation by movie is insufficient, and using actual ball and pipes in person is expected to be most effective to explain programming behavior.

While answers to question (i) was not improved. On explanation 3, we showed an example which explains that type A is improper to keep a bell ringing. This might affect correct answer to question (i).

**Factors affecting the effect of visualization**; We analyzed the relationship between the answers of the test, and the feeling of inadequacy and the knowledge of the participants. Spearman's rank correlation coefficient was used for the analysis. As explained in Section III, the participants answered the questions three times. We denote answers of them as 1-i,1-ii, 2-i, 2-ii, 3-i, and 3-ii respectively. The Arabic numbers denote types of explanations shown in Section III. When the answers were correct, we set the value as 1 and otherwise, set as 0.

Table 2 shows the correlation coefficient and p-values of the answers and the knowledge. The relationships between feeling of inadequacy and most results of the tests (2-i, 2-ii, 3-i, and 3-

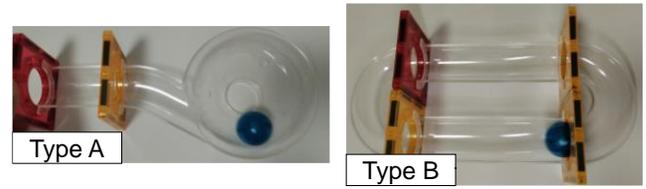

Fig. 2  **Two types of visualized program loop**

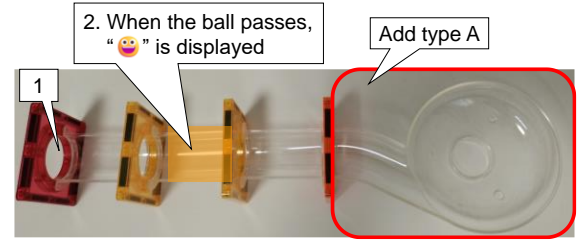

Fig. 3  **Addding program loop to a program on program (i)**

TABLE I.     RATE AND NUMBER OF CORRECT ANSWERS [4]

|  | Explanation 1 | Explanation 2 | Explanation 3 |
|---|---|---|---|
| **Question (i)** | 57% (4) | 57% (4) | 43% (3) |
| **Question (ii)** | 43% (3) | 57% (4) | 71% (5) |

TABLE II.    RELATIONSHIP BETWEEN THE KNOWLEDGE AND ANSWERS

|  |  | 1-i | 1-ii | 2-i | 2-ii | 3-i | 3-ii |
|---|---|---|---|---|---|---|---|
| **I2** | ρ | 0.45 | 0.60 | 0.60 | 0.90 | 0.30 | 0.49 |
|  | p-value | 0.31 | 0.16 | 0.16 | 0.01 | 0.51 | 0.26 |
| **I3** | ρ | 0.51 | 0.44 | 0.36 | 0.87 | 0.07 | 0.64 |
|  | p-value | 0.24 | 0.33 | 0.42 | 0.01 | 0.88 | 0.12 |

ii) were strong. Therefore, there is a room for improvement when using the visualization method, although the percentage of correct answer to question (ii) was high, when explanations 3 was used.

The relationship between I1 and I2, 3 was relatively strong. The former one of ρ was -0.75 and the latter one was -0.74. This suggests that feeling of inadequacy related to the knowledge. While the knowledge did not related to the comprehension tests strongly, except for 2-i and 2-ii. Therefore, the correctness of the tests did not depend on the knowledge strongly. This suggests that the visualization method is effective to understand program behavior to some extent.

## References

[1] T. Bell, M. Fellows, and I. Witten, Computer Science Unplugged: Off-line Activities and Games for All Ages, Computer Science Unplugged, 1998.

[2] W. Huang, and C. Looi, "A critical review of literature on "unplugged" pedagogies in K-12 computer science and computational thinking education," Computer Science Education, vol.31, no.1, pp.83-111, 2021.

[3] S. Jinnouchi and M. Tsunoda, "Visualizing Program Behavior with a Ball and Pipes for Computer Science Unplugged," In Proc. of Asia-Pacific Software Engineering Conference (APSEC), pp.663-664, 2023.

[4] S. Jinnouchi and M. Tsunoda, "Visualizing Program Behavior with a Ball and Pipes for Computer Science Unplugged," In Proc. of JSSST Workshop on Foundation of Software Engineering (FOSE), 2024. (to appear)

[5] KitWell, Magbuild, https://www.amazon.co.jp/dp/B08LPGVDMH?language=en_US